\documentclass[letter]{aa} 

\usepackage{txfonts}
\usepackage{graphicx,url,twoopt,natbib}
\usepackage[breaklinks=true]{hyperref} 
\bibpunct{(}{)}{;}{a}{}{,} 
\makeatletter
\newcommandtwoopt{\citeads}[3][][]{\href{http://adsabs.harvard.edu/abs/#3}%
    {\def\hyper@linkstart##1##2{}%
\let\hyper@linkend\@empty\citealp[#1][#2]{#3}}}
\newcommandtwoopt{\citepads}[3][][]{\href{http://adsabs.harvard.edu/abs/#3}%
{\def\hyper@linkstart##1##2{}%
\let\hyper@linkend\@empty\citep[#1][#2]{#3}}}
\newcommandtwoopt{\citetads}[3][][]{\href{http://adsabs.harvard.edu/abs/#3}%
{\def\hyper@linkstart##1##2{}%
\let\hyper@linkend\@empty\citet[#1][#2]{#3}}}
\newcommandtwoopt{\citeyearads}[3][][]%
{\href{http://adsabs.harvard.edu/abs/#3}
{\def\hyper@linkstart##1##2{}%
\let\hyper@linkend\@empty\citeyear[#1][#2]{#3}}}
\makeatother

\begin{document}

   \title{Formation of single-moon systems around gas giants}

   \author{Yuri I. Fujii\inst{1}\inst{2}
          \and
          Masahiro Ogihara\inst{3}}

   \institute{Institute for Advanced Research, Nagoya University, 
                Furo-cho, Chikusa-ku, Nagoya, Aichi, 464-8601, Japan\\
              \email{yuri.f@nagoya-u.jp}
         \and
            Department of Physics, Nagoya University, Furo-cho, Chikusa-ku, Nagoya, Aichi, 464-8602, Japan
         \and
            Division of Science, National Astronomical Observatory of Japan,
            2-21-1, Osawa, Mitaka, 181-8588 Tokyo, Japan\\
             }

   \date{Received ; accepted }

 
  \abstract
   {Several mechanisms have been proposed to explain the formation process of satellite systems,
       and relatively large moons are thought to be born in circumplanetary disks. Making a single-moon
       system is known to be more difficult than multiple-moon or moonless systems.
   }
   {We aim to find a way to form a system with a single large moon, such as Titan around Saturn. 
       We examine the orbital migration of moons, which change their direction and speed depending 
    on the properties of circumplanetary disks.
   }
   {We modeled dissipating circumplanetary disks with taking the effect of temperature structures
       into account and calculated the orbital evolution of Titan-mass satellites in the 
       final evolution stage of various circumplanetary disks.
       We also performed $N$-body simulations of systems that initially had multiple satellites 
       to see whether single-moon systems remained at the end.
   }
   {The radial slope of the disk-temperature structure characterized by the dust opacity
       produces a patch of orbits in which the Titan-mass moons cease inward migration and even migrate 
       outward in a certain range of the disk viscosity. The patch assists
       moons initially located in the outer orbits to remain in the disk, while 
       those in the inner orbits fall onto the planet.
   }
   {We demonstrate for the first time that systems can form that have only one large moon around giant planet. 
   Our $N$-body simulations suggest satellite formation was not efficient in the outer radii of 
   circumplanetary disks. 
   }

   \keywords{Planets and satellites: formation -- Planets and satellites: individual: Titan --
       Planets and satellites: dynamical evolution and stability --
       Planet-disk interactions --Planets and satellites: gaseous planets 
               }

   \maketitle
%

\section{Introduction}
               \label{sec:intro}
Satellite formation around gas giant planets has been widely discussed in the context of the Galilean moons. 
The small moons around gas/ice giants in our solar system can also be explained by a tidally 
spreading solid disk scenario \citep{cha10, cri12, hyo15} when it is possible to provide the enough 
material for the initial solid disk.
For large moons, 
such as Io, Europa, Ganymede, and Callisto around Jupiter, and Titan orbiting Saturn,
however, it is favorable to have a gaseous circumplanetary disk (CPD) for their formation. 
Models for CPDs were proposed and the formation of Galilean moons in the disks 
has been discussed \citep{can02, can06, mos03a, mos03b, est06, sas10, ogi12, mig16, fuj17, 
cil18, shi19, ara19}. 
Saving satellites from inward migration and configuring a system in a Laplace resonance are 
current topics of strong interest.
We can find from these studies that it is difficult to form only one large moon. Satellite systems
tend to hold multiple large moons or lose all of these moons at the end of the simulations.
Although \citet{sas10} demonstrated that single-moon systems form as a result of population
synthesis calculations, it is likely that their method for generating
new moon seeds is not appropriate for moon formation (see Section 5 of \citet{ogi12}). 

We aim to find a way to form a system with only one large moon in a CPD.
The feasibility of satellitesimal formation has been examined by \citet{shi17}
and delivery of solid matereal to a CPD has been discussed by 
\citet{fuj12, tan14, sue17, ron18}.
We focus on the later stage of 
the satellite formation  and investigate the orbital evolution of the moons
in a dissipating CPD to determine the final appearance of the system.

\section{Disk model}
               \label{sec:disk}

   \begin{figure}
   \centering
   \includegraphics[scale=0.35]{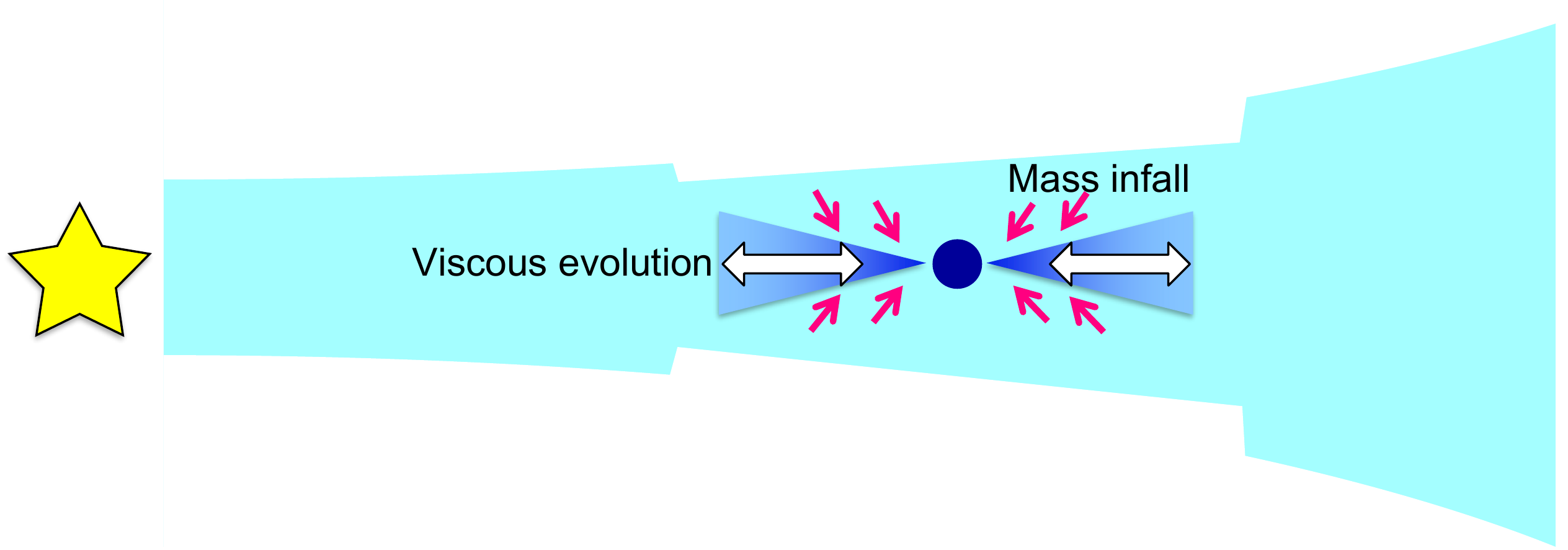}
         \caption{Schematic picture of how to determine the surface density; viscous evolution
         of the CPD balances with the mass infall from the PPD.
         }
         \label{fig:infall}
    \end{figure}
%
   \begin{figure}
   \centering
   \includegraphics[scale=1.0]{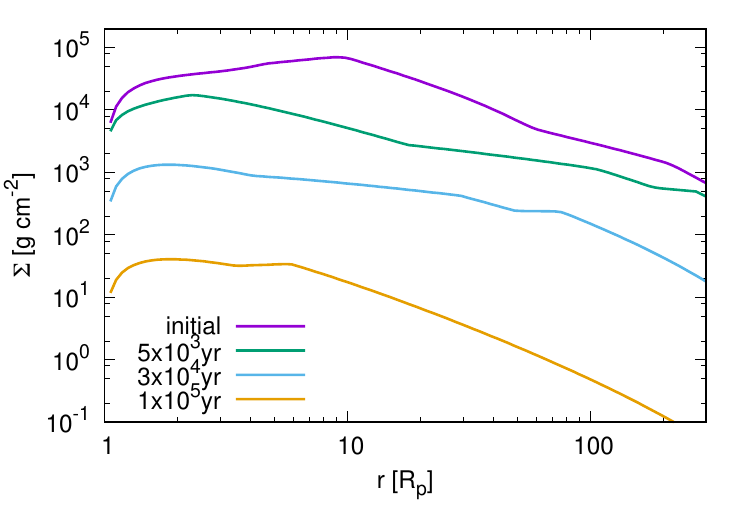}
   \includegraphics[scale=1.0]{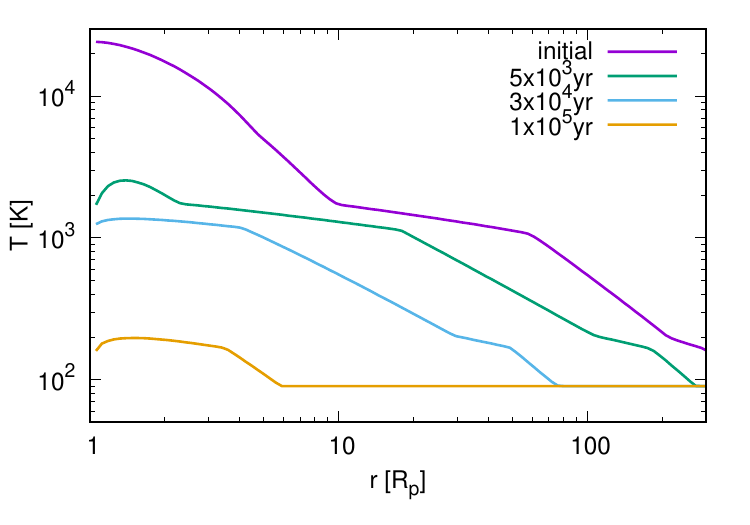}
   \caption{Time evolution of the surface density (top) and temperature (bottom) 
       structure of the dissipating CPD with $\alpha=10^{-4}$.
       The time is measured from when the mass infall has terminated.
   }
         \label{fig:disk}
    \end{figure}
We derived 1D models of CPDs considering the effect of opacities that can change 
not only the temperature but also the surface density structures of the disks. 
The surface density, $\Sigma$, is determined by the balance of the mass infall from the parental 
protoplanetary disk (PPD) and the viscous evolution of the CPD as described in the Fig. \ref{fig:infall},
and its time evolution can be calculated with the following diffusion equation:
\begin{eqnarray}
    \frac{\partial \Sigma}{\partial t}
    =  \frac{1}{r}\frac{\partial }{\partial r}
    \left[ 3r^{1/2}\frac{\partial}{\partial r}
            \left( r^{1/2}\nu\Sigma \right)\right]
            + f,
    \label{dSdt}
\end{eqnarray}
where 
$r$ is the radial distance from the planet, $\nu$ is the  
kinematic viscosity coefficient, and $f$ is the mass flux from the PPD
that has the Keplerian angular momentum of the corresponding radii of the CPD.
\citet{tan12} show $f \propto r^{-1}$ and we adopt 
$f= 5.5\!\times\!10^{-5} (r/R_{\rm p})^{\!-1} {\rm g\,cm^{-2}\,s^{-1}}$,
where the radius of the planet $R_{\rm p}$ is the Saturnian radius,
until the disk starts to dissipate.
We assume the planet is located at 9.5 AU of the minimum mass solar nebula (MMSN) \citep{hay81} 
and give $f$ only within $r_{\rm infall}=45R_{\rm p}$ (0.04 Hill radii) in this work 
\citep[see][for the details]{fuj17}.
The final configuration of the system is not sensitive to the choice of $r_{\rm infall}$,
and changing it to $30R_{\rm p}$ or $60R_{\rm p}$ does not change the result. 
We set the size of the CPD as $\sim 0.7$ Hill radii.

The disk midplane temperature, $T_{c}$, is calculated by  
\begin{eqnarray}
    \frac{\partial T_c}{\partial t}
    = \frac{2\left( Q_+ - Q_- \right)}{c_p \Sigma}
    - v_r\frac{\partial T_c}{\partial r},
    \label{dTdt}
\end{eqnarray}
where $Q_+ = (9/8)\nu\Sigma\Omega_{\rm K}^2$ is the viscous heating,
$Q_- = \sigma \left(3/8\kappa \Sigma \right)T_c^4$ denotes the
radiative cooling, $\sigma$ is the Stefan-Boltzmann constant,
$v_{r}$ is the radial velocity, and $\kappa$ is the opacity \citep{can93, arm01}.
The opacities employed are provided by \citet{bel94}. We also modeled CPDs 
by assuming an order of magnitude smaller dust/ice opacities (see Sect. \ref{sec:con}). 
We set the minimum temperature of the disk as {$T_{\rm min} = 90$ K. 

The origin of the angular-momentum transport in a CPD is uncertain. 
The magnetorotational instability (MRI), which is believed to be one of the sources of 
the turbulent viscosity in a PPD, is not very effective in a CPD except for the vicinity
of the planet \citep{fuj11, fuj14, tur14, kei14}. 
Thus, we assume the mass accretion is gentle in a CPD.
We leave the origin of the viscosity unspecified and adopt $\nu=\alpha c_{\rm s} h$ with 
viscous parameter $\alpha$ of \citet{sha73}, 
sound speed $c_{\rm s}$ and pressure scale height of the disk $h$.

With a fixed viscous parameter, the CPD settles into a steady state as 
long as it has the constant infall. 
For example, the purple lines in Fig. \ref{fig:disk} show the steady-state 
disk structure for $\alpha=10^{-4}$. 
Unlike \citet{fuj17}, we do not introduce a scaling factor to modify the  
local surface density of the PPD because this parameter is coupled with the 
viscous parameter when we derive the steady state. The value of $f$ corresponds to 
the surface density of the PPD that is reduced to 10\% of the MMSN. 
In order to investigate the final stage of the satellite formation, 
we calculate the time evolution of 
CPDs after the mass inflow from the PPD has terminated
($f$ is set to be zero). 
The manner of the disk dissipation in the case with $\alpha=10^{-4}$
is described in Fig. \ref{fig:disk}.
We assume the timescale of halting the infall is small enough.

\section{Orbital evolution of moons}
               \label{sec:mig}

   \begin{figure}
   \centering
   \includegraphics[scale=0.32]{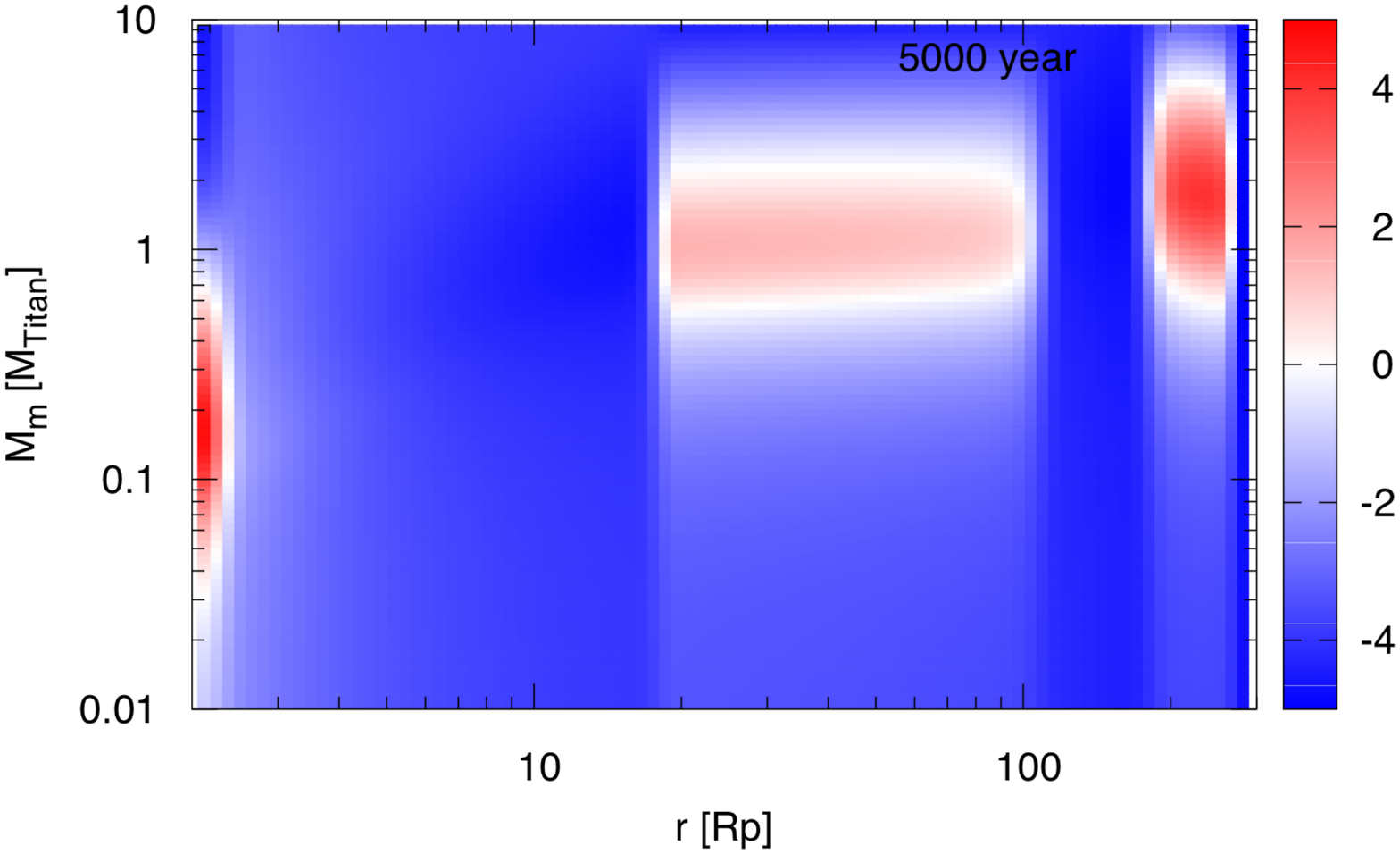}
   \caption{Migration map for a disk with $\alpha=10^{-4}$ at 5000 years after the
       dissipation has started (corresponding to the green lines in Fig. \ref{fig:disk}). 
       The color bar shows the value of 
   $\beta/(1+0.04K)$. The direction of a migration is inward/outward in the blue/red region.
   }
               \label{fig:betamap}
    \end{figure}
%
   \begin{figure}
   \centering
   \includegraphics[scale=1.0]{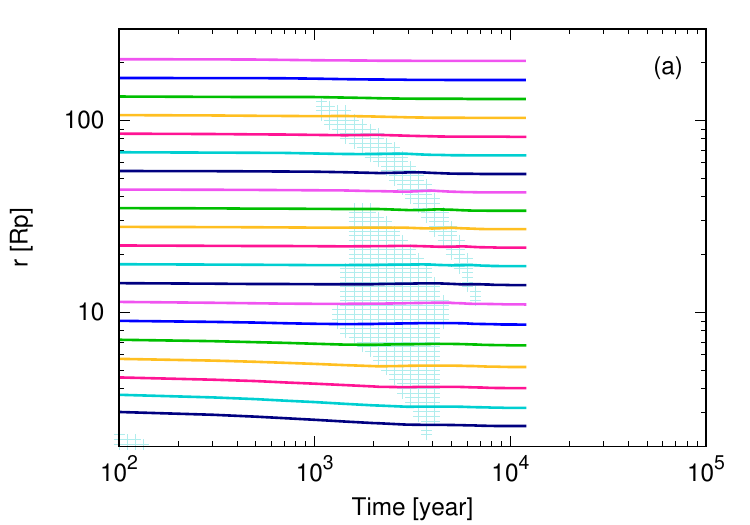}
   \includegraphics[scale=1.0]{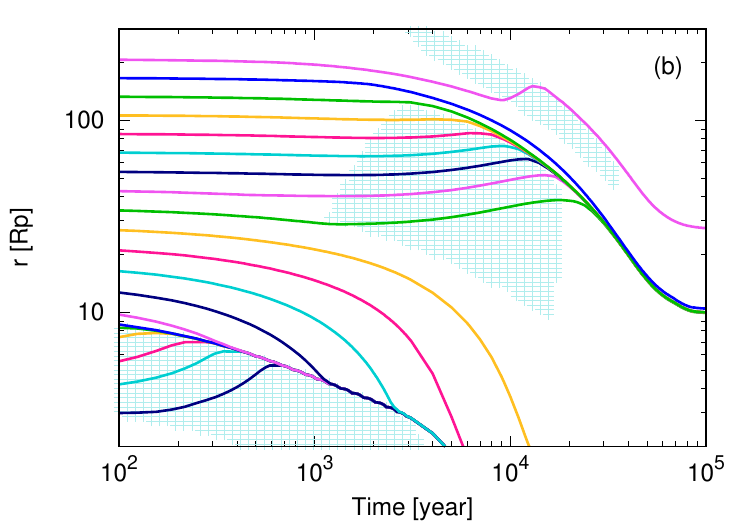}
   \includegraphics[scale=1.0]{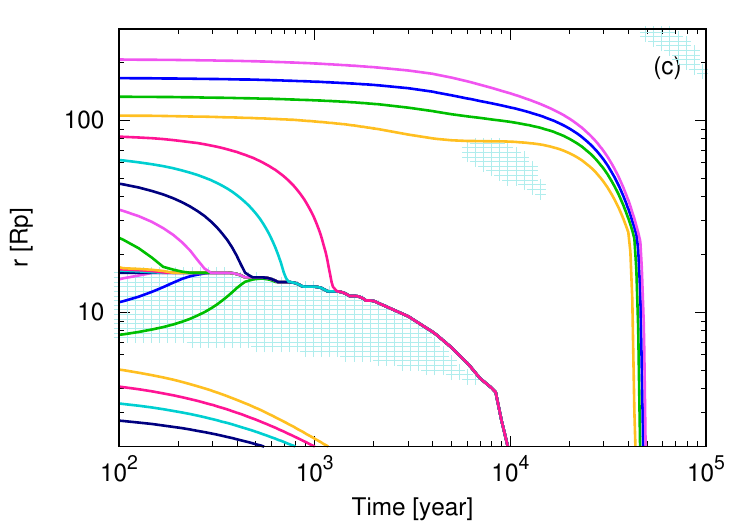}
   \caption{Orbital evolution of individual moons since the disk dissipation started
       with various initial locations in the 
               case where $\alpha=10^{-3}$, $\alpha=10^{-4}$ and $\alpha=10^{-5}$ from top to bottom. The vertical axis is normalized by the planet radius. 
               The shaded areas show when and where $\beta>0$ for Titan mass moons, which
               correspond to the red regions in Fig. \ref{fig:betamap}. 
               The mutual interactions of satellites are not considered in these plots.
           }
               \label{fig:orbit}
    \end{figure}
We calculated the orbital evolution of satellites in the dissipating CPDs from 
Sect. \ref{sec:disk} to determine the configuration of the system. 
The orbital radius of a moon is determined by
\begin{equation} \frac{d r}{d t} = \beta~(1 + 0.04K)^{-1}\frac{M_m}{M_{p}}
    \frac{\Sigma r^2}{M_{p}}
    \left(\frac{r \Omega_{K}}{c_{s}}\right)^{2}r \Omega_{\rm K},
    \label{dadt}
\end{equation}
where $M_{\rm m}$ and $M_{\rm p}$ are the mass of the moon and planet, respectively, 
$\beta$ is a migration parameter that determines the
direction and speed of the migration \citep{paa11,kre12,ogi15}.
The factor $(1 + 0.04K)$ connects type I and type II migration using
$K=(M_m/M_p)^2(h/r)^{-5}\alpha^{-1}$ \citep{kan15, kan18a, ogi}.  
The migration parameter depends on the radial dependence of the surface density and 
the temperature of the disk. When $\beta$ is positive, the moon migrates outward. 
Otherwise, the direction of the satellite migration is toward the planet.
We used the formula of \citet{paa11} for $\beta$\footnote{See the Appendix A of \citet{bit11}
about the factor 4 of the thermal diffusivity.}, which is also summarized in \citet{ogi15}.
The migration parameter also depends on the 
eccentricity $e$ and inclination $i$ of an orbit \citep{bit10}, and we 
assumed $e=i=0$ in this section. Those dependences \citep{cre08} are, however, taken into account
in the Sect. \ref{sec:N}. 
The map of $\beta$ for satellites of $0.01 - 10$ Titan mass in the disk with $\alpha=10^{-4}$ 
is shown in Fig. \ref{fig:betamap}.
The red patches in the figure indicate the parameter space with outward migration. 
The patch located at $r=\sim20-100R_{\rm p}$ is produced by the shallow 
and steep radial dependence of $\Sigma$ and $T$ characterized by the dust opacity. 
The outer patch located at $r > 100 R_{\rm p}$ reflects the opacity due to ice.
We can see that Titan-mass satellites captured in these patches escape from inward migration. 

Fig. \ref{fig:orbit} demonstrates the orbital evolution of moons with Titan mass.
We note that the plots are meant for showcasing the individual evolution of each of the moons in the disks,
and therefore, the gravitational interactions with other moons are not considered in the calculation.
With $\alpha=10^{-3}$, as shown in Fig. \ref{fig:orbit}(a), satellites tend to survive until 
the disk disappears; thus, if there are multiple moons at the beginning, the system would keep 
the moons for a longer time period. This is partially because of the saturation of the corotation torque
and also because of the quick dissipation of the disk.
On the other hand, all the moons are lost as a consequence of the inward migration in the case of 
$\alpha=10^{-5}$ (Fig. \ref{fig:orbit}(c)). In this case, the timescales of the migration in the inner orbits are 
much shorter than in the case of $\alpha=10^{-3}$ and it takes longer until the disk disappears. 
The most interesting parameter is $\alpha=10^{-4}$, with which 
the moons in the inner orbits fall onto the central planet but others in the outer orbits
remain in the disk as shown in Fig. \ref{fig:orbit}(b).
The patch of $\beta>0$ created by the dust opacity prevent moons from moving toward the planet,
while inner orbits are cleared. Whether the surviving satellites remain in a couple of tens 
Saturnian radii is determined by the timing of the disk dissipation. 

How many satellites can form and exist at one time depends not only
the CPD structure but also the manner and amount of the solid supply. 
We found that even if many moons form in the disk, most of them would be lost and only
a couple of them remain when the CPD has been dissipated. Through this mechanism,  
we can possibly form a single-moon system unless the satellite formation in the large 
orbit is very efficient. 
In order to further investigate the possibility, we carry out $N$-body simulations in the next section.
Although it might be challenging to create many satellites in a CPD, we insert 
relatively large number of moons as our goal is to make a single-moon system.

\section{\textit{N}-body simulations}
               \label{sec:N}

In the calculations shown in the previous section, the mutual gravitational
interaction between moons is not considered. In this section, we perform
simulations that include the \textit{N}-body interaction and examine whether
single-moon systems can form with the initial setting supposing optimistically 
many moons can form before the CPD starts to dissipate. We do not include moons that 
form during the run because we assume there is no supply of solid when gas inflow
from the PPD has finished.
   \begin{figure}
   \centering
   \includegraphics[scale=0.7]{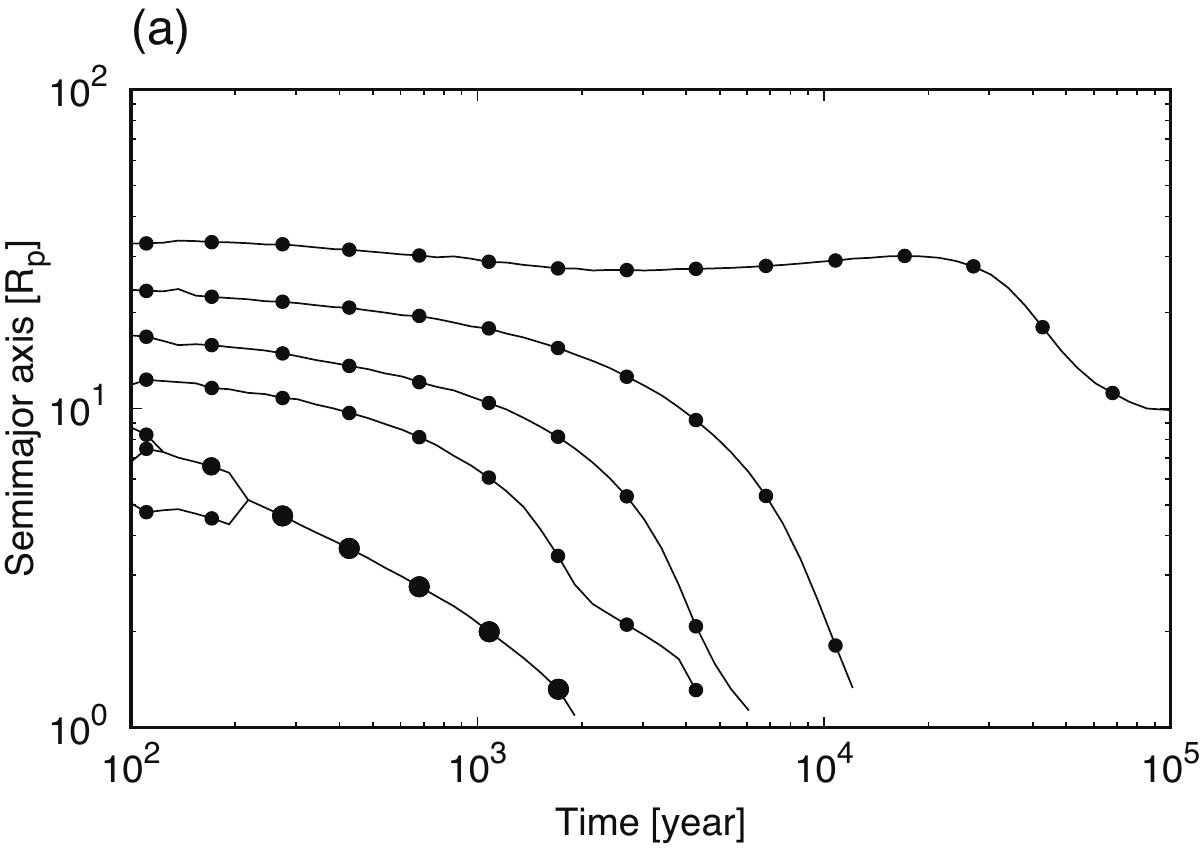}
   \includegraphics[scale=0.7]{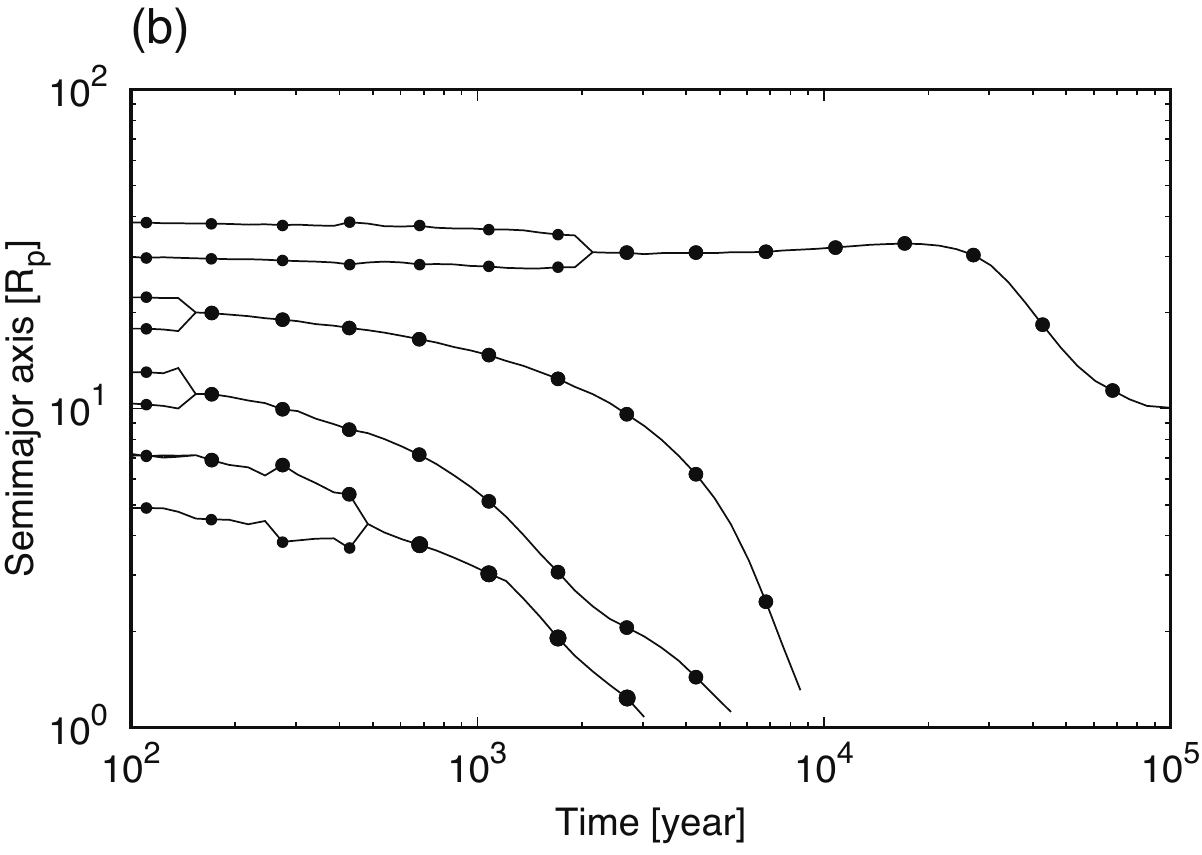}
   \caption{Results of the N-body simulations. 
       (a) The case with initially seven moons of the Titan mass. 
       (b) The case with nine moons with 0.5 Titan masses.
       We terminated the simulations when the disk gas has mostly dissipated 
       and no more satellite migration was expected.
   }
               \label{fig:Nbody}
    \end{figure}
Fig. \ref{fig:Nbody} shows the time evolution of the semimajor axis.
In Fig. \ref{fig:Nbody}(a),
seven moons with one Titan mass are initially placed between 
$5 \,R_{\rm p}$ and $40 \,R_{\rm p}$ with separations of 6 mutual Hill
radii. As seen in Fig. \ref{fig:orbit}, while inner moons migrate inward and fall
onto the planet, a moon in an outer orbit undergoes outward migration
and avoids falling onto the planet. Finally, the moon remains in the
outer orbit and a single-moon system forms.
Fig. \ref{fig:Nbody}(b) shows simulation result that starts with nine moons
with 0.5 Titan masses. We also see that inner moons fall onto the planet
and a single moon with a Titan mass remains.

Although multiple moons remain in previous \textit{N}-body simulations
(e.g., Canup \& Ward 2006; Ogihara \& Ida 2012), we first demonstrate
that single-moon systems can form under some conditions.
As already stated above, the number of remaining moons depends on the
property of CPD and the solid distribution. For example, when we start
simulations with moons that extend beyond $40 \,R_{\rm p}$, the number
of final moons tends to increase. This is consistent with the
hypothesis that regular moons grow in the inner region, which
justifies our assumption of initial distribution of moons.

\section{Discussions and conclusions}
               \label{sec:con}
We demonstrated the qualitative pathway to establish
a single-moon system:
   \begin{enumerate}
      \item Moons in the outer orbits are captured in the patch with outward migration.
      \item Inner orbits are cleared by type I migration.
      \item The disk disappears before the final moon falls onto the planet.
   \end{enumerate}

We find that there is a favorable value of the viscous parameter to form a system
that has a single Titan-mass satellite; this value is $\alpha=10^{-4}$ in our settings. 
With $\alpha=10^{-3}$, more than one of the moons tend to survive if they have formed.
In the case of $\alpha=10^{-5}$, all the moons would be lost unless the disk dissipation 
has accelerated.
As shown in Fig. \ref{fig:betamap},
the direction of migration changes with the mass of moons, thus the conditions to 
be a single-moon system differs in the cases for other masses.
Other than the migration parameter, the timescale for the disk dissipation is 
also an important factor. 

We employed the $\alpha$-disk model for simplicity, and assumed the value of 
$\alpha$ as a constant both in time and space. Even in a dissipating CPD, the MRI may not contribute \citep[see Fig. 6 of][]{fuj14}, however, other mechanisms may
vary the dynamical evolution of the disk. In such a case, the condition to obtain a single-moon
system and the location of the moon would be modified.

We also calculated cases with an order of magnitude smaller dust and ice opacities 
compared with those of \citet{bel94}. Of course, the disk surface-density and temperature structures
are modified but the tendency of the orbital evolution of the moons in the disks
were similar to the case with the original opacity. In the lower opacity case,
the disk should dissipate a little too quickly.

\begin{acknowledgements}
      We thank the anonymous referee for useful comments.
      YIF was supported by the JSPS KAKENHI Grant Number JP18K13604 and
      Start-up grant of Building of Consortia for the 
      Development of Human Resources in Science and Technology from 
      Ministry of Education,
      Culture, Sports, Science, and Technology (MEXT).
      Numerical computations were in part carried out on PC cluster at 
      Center for Computational Astrophysics, National Astronomical Observatory of Japan.
\end{acknowledgements}

%
%

\bibliographystyle{aa} 
\bibliography{manuscript} 

\end{document}